\documentclass[letterpaper]{article}
\usepackage{aaai}
\usepackage{times}
\usepackage{helvet}
\usepackage{courier}

\usepackage{graphicx}
\usepackage{algorithm}
\usepackage{algorithmic}
\usepackage{amsmath,amssymb,amsfonts}
\usepackage{textcomp}
\usepackage{xcolor}
\usepackage{multirow}
\usepackage{flushend}

\begin{document}
%
\title{Deep Hashing for Signed Social Network Embedding}
\author{
Jia-Nan Guo, Xian-Ling Mao, Xiao-Jian Jiang, Ying-Xiang Sun, Wei Wei and He-Yan Huang\\
Beijing Institute of Technology, Beijing, China\\
Huazhong University of Science and Technology, China\\
\{guojn97, jiangix321\}@gmail.com\\
\{maoxl, bitcssyx, hhy63\}@bit.edu.cn\\
Weiw@hust.edu.cn\\
}
\maketitle
\begin{abstract}
Network embedding is a promising way of network representation, facilitating many signed social network processing and analysis tasks such as link prediction and node classification. Recently, feature hashing has been adopted in several existing embedding algorithms to improve the efficiency, which has obtained a great success. However, the existing feature hashing based embedding algorithms only consider the positive links in signed social networks. Intuitively, negative links can also help improve the performance. Thus, in this paper, we propose a novel deep hashing method for signed social network embedding by considering simultaneously positive and negative links. Extensive experiments show that the proposed method performs better than several state-of-the-art baselines through link prediction task over two real-world signed social networks.

\end{abstract}

\section{Introduction}
As the availability of large-scale social media network increases, the traditional network representation methods have limited performance due to their sparsity. In order to address the problem, network embedding is developed to learn a continuous low-dimensional representation for each node of a network, which preserves both properties and structure information of the node. Currently, a significant amount of progresses have been made toward network embedding, including DeepWalk \cite{perozzi2014deepwalk}, Node2Vec \cite{grover2016node2vec}, SiNE \cite{wang2017signed} and SNE \cite{yuan2017sne}. This kind of network representation method has greatly facilitated advanced network analytic tasks in both time and space, such as link prediction \cite{liben2007link}, \cite{ou2016asymmetric} and node classification \cite{sen2008collective}.

Generally, the binary codes can further facilitate to represent and search of massive data. Therefore, several feature hashing based network embedding methods have been proposed to further improve efficiency of similarity search in embedding space and reduce storage of nodes representation, such as Node2Hash \cite{wang2018feature} and NetHash \cite{wu2018efficient}.  Node2Hash uses the encoder-decoder framework \cite{hamilton2017representation}, where the encoder is used to map the proximity of nodes into a feature space, and the decoder is used to generate node embedding through feature hashing. NetHash denotes each node of a network as a shallow rooted tree, and then adopts LSH \cite{indyk1998approximate} to embed it from bottom to top. Both of them have improved the performance of KNN search in the area of network embedding as hash codes are storage efficient and permit exact sub-linear KNN search.

To the best of our knowledge, the existing feature hashing based embedding algorithms only consider the positive links in signed social networks. However, signed social networks generally have both positive and negative links. For example, Epinions \cite{wang2017signed} allows users to mask other users as trust or distrust on product review, and Slashdot \cite{wang2017signed}, \cite{yuan2017sne} allows users to specify other users as friends or foes.

Intuitively, the negative links in signed social networks can help improve the performance of feature hashing based social network processing and analysis tasks. Recent researches also confirm the viewpoint in other areas. For example, \cite{leskovec2010predicting} finds that a small number of negative links can significantly improve the performance of positive link prediction in online social networks. Moreover, the aforementioned problem cannot be solved by simply extending the algorithm used for unsigned social networks, because some researches suggest that homophily effects and social influence for unsigned network may not be applicable to signed network \cite{tang2014distrust}. 

Therefore, we propose a novel deep Hashing method for Signed social Network Embedding by considering simultaneously positive and negative links, named HSNE. Specifically, the proposed method consists of three key components: (1) Network feature learning; (2) Hash code learning and (3) Loss function. The first component is designed to learn both properties and structure features from a signed social network. The second component is used to transform network features to hash codes by a fully connected layer. Finally, in order to incorporate the structure and property-level constraints of a signed social network into deep model, a triplet loss function is designed to preserve the features of positive and negative links.

The main contributions of this paper are as follows:
\begin{itemize}
\item We introduce deep hashing into network embedding to further improve efficiency of similarity search in embedding space and reduce storage of nodes representation.
\item We propose a novel deep hashing method for signed social network embedding, which can consider simultaneously positive and negative links.
\item Extensive experiments show the proposed method performs better than several state-of-the-art baselines, which only consider positive links in signed social networks, through link prediction task over two real-world signed social networks.
\end{itemize}

\section{Related Work}

\subsection{Network Embedding}

Network embedding aims to learn a low-dimensional representation for each node of a network. To transform networks from original network space to embedding space, lots of models have been proposed. The commonly used frameworks include matrix factorization, random walk, deep neural network and their variations. The matrix factorization based methods are to learn a low-rank space for adjacency matrix including Singular Value Decomposition based method \cite{ou2016asymmetric} and Non-negative matrix factorization based method \cite{wang2017community}. The random walk based models are exploited to generate random paths over a network, and the node neighborhood can be identified by co-occurence rate. There are some representative methods, such as DeepWalk \cite{perozzi2014deepwalk} and Node2vec \cite{grover2016node2vec}. The deep neural network methods provide end-to-end solutions including SDNE \cite{wang2016structural} and SDAE \cite{cao2016deep}. 

Recently, several algorithms have been proposed to consider simultaneously positive and negative links in signed social network, such as SiNE \cite{wang2017signed}, SNE \cite{yuan2017sne}. Among all the existing related network embedding method, SiNE is the closest to our work. Nevertheless, their work is mapping nodes into 
continuous space, whose time complexity over finding similar nodes in embedding space is $\mathcal{O}(n^2)$. As the size of network growing larger and larger, similarity search in the embedding space becomes impractical. In contrast, we combine SiNE and deep hashing to map nodes into Hamming space to improve efficiency.

\subsection{Deep Hashing}

Deep hashing methods have been proposed to simultaneously learn features and hash codes using deep neural networks. It has been shown superior performance over traditional hashing methods. The existing deep hashing methods can be further divided into two categories \cite{wang2018survey}: unsupervised methods and supervised methods. The unsupervised deep hashing methods either are usually based on the reconstruction loss or preserve the similarity between rotated image, such as Semantic hashing \cite{salakhutdinov2009semantic}, deep auto-encoder hashing \cite{hinton2011transforming} and Semantic structure-based hashing \cite{zhao2015deep}.  The supervised deep hashing methods try to utilize label information to learn hash codes. Most deep hashing methods are given supervised information in the form of pairwise labels or triplet labels. For example, DPSH \cite{li2015feature} proposes two deep models with shared parameters to maximize the likelihood of the given pairwise labels. \cite{wang2016deep} proposes three deep models with shared parameters to maximize the likelihood of the given triplet labels. Recent research suggests that triplet labels inherently contain richer information than pairwise labels \cite{norouzi2012hamming}. 

To the best of our knowledge, currently, there have been two hashing based network embedding methods, Node2Hash \cite{wang2018feature} and NetHash \cite{wu2018efficient}. However, the two existing algorithms just use shallow hashing methods, and only consider positive links in signed social network. In contrast, we propose a novel deep hashing method for signed social network embedding, which considers simultaneously positive and negative links to improve effectiveness. 

\begin{figure*}[htbp]
\centering
\includegraphics[width=14cm,height=8cm]{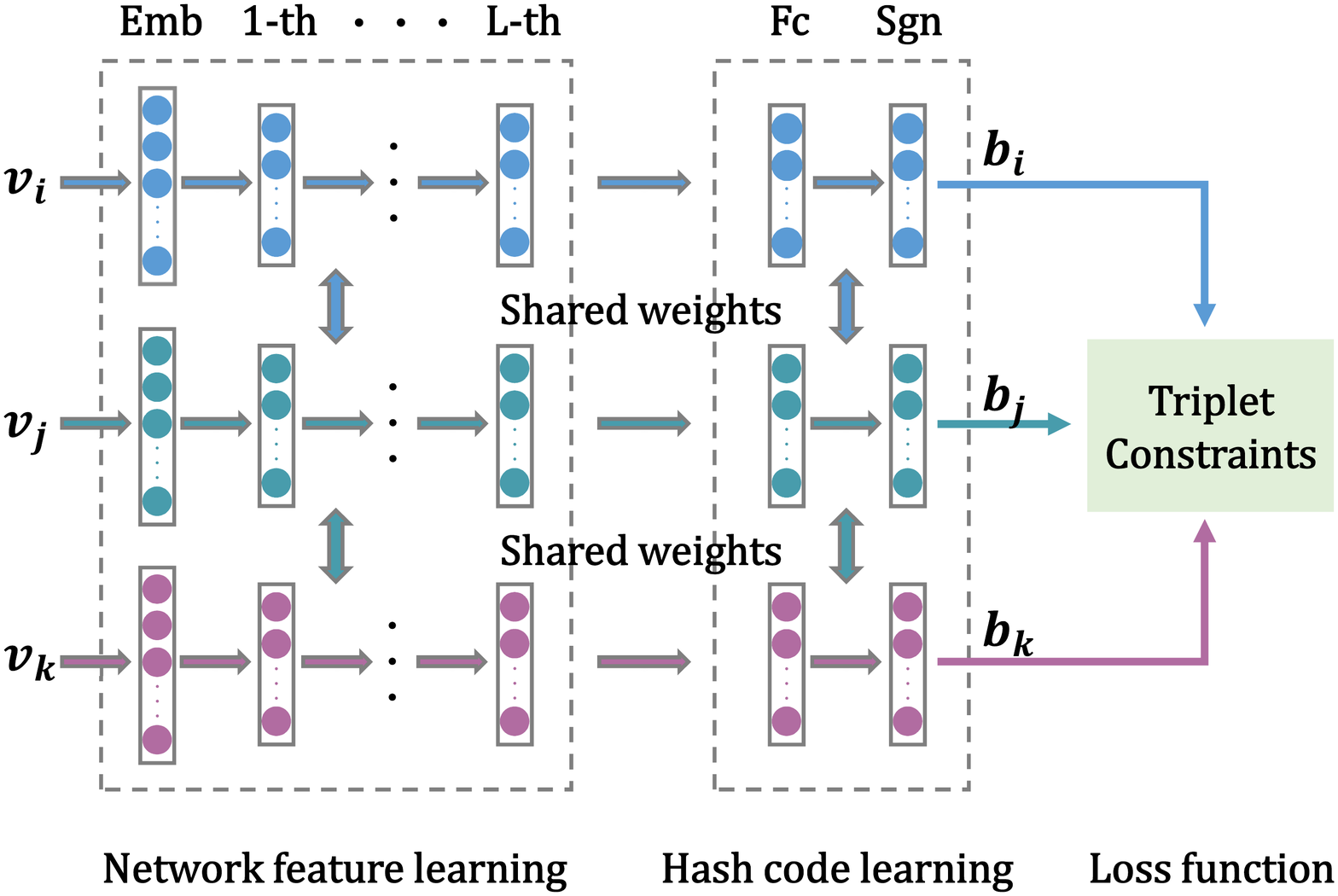}
\caption{Overview of the proposed deep hashing based method (HSNE), which contains three key components: (1) Network feature learning; (2) Hash code learning and (3) Loss function.}
\label{fig1}
\end{figure*}

\section{Approach}
Before introducing the details of the proposed approach, the notations used in this paper will be introduced first. Boldface uppercase letters like $\mathbf{W}$ denote matrices, and vectors are written in boldface lowercase letters like $\mathbf{w}$.  Capital letters in calligraphic math font like $\mathcal{V}$ are used denote sets, and $|\mathcal{V}|$ is the cardinality of $\mathcal{V}$. In this way, a signed network can be expressed as $\mathcal{G} = \{\mathcal{V}, \mathcal{E}\}$ where $\mathcal{V} = \{v_1, v_2, ..., v_M\}$ is a set of M nodes and $\mathcal{E} = \{e_{ij}\}$, $v_i,v_j \in \mathcal{V}$ is a set of links.

\subsection{Problem Definition}
Construct a training set of triplets, $\mathcal{T} = \{(v_i, v_j, v_k)\}$, according to the positive and negative links of a signed social network, where $e_{ij} = 1, e_{jk} = -1, e_{ij}, e_{jk} \in \mathcal E $. In signed social network, it's well recognized that positive links express the similarity and negative links express the dissimilarity. Thus, any triplet data point $(v_i, v_j, v_k) \in \mathcal{T}$ has relative similarity label or triplet label \cite{norouzi2012hamming}, where the pair $(v_i, v_j)$ is more similar than the pair $(v_i, v_k)$. The goal of supervised hashing is to learn binary hash codes for all the nodes from $\mathcal{T}$, and the learned hash codes try to preserve the relative similarity. More specifically, if we use the $\mathcal{B} = \{\mathbf{b_i}\}_{i=1}^m , \mathbf{b_i} \in \{-1, +1\}^{d}$ to denote the learned binary hash codes for $\mathcal{V}$, the Hamming distance $\| \mathbf{b_i} - \mathbf{b_j} \|_H$ should be more closer than $\| \mathbf{b_i} - \mathbf{b_k} \|_H$. Here, $d$ denotes the length of binary code.

\subsection{The proposed method}
The existing feature hashing based embedding algorithms only consider the positive links in signed social networks. Thus, in this paper, we propose a novel deep hashing method for signed social network embedding by considering simultaneously positive and negative links. As shown in Fig.~\ref{fig1}, our method has three key components: (1) Network feature learning; (2) Hash code learning and (3) Loss function. We will introduce the three components in detail in the following parts. 

{\bf Network feature learning}: This component is designed to employ a deep neural network to learn features from signed social network. We adopt an embedding layer and $L=3$ fully connected layers for this component. The embedding layer is designed to embed nodes into low-dimension vectors, which has $d_0$ units. Three fully connected layers are designed to preserve the positive and negative links of a signed social network, and have $d_i$ units for i-th layer. Besides, we choose the hyperbolic tangent $tanh$ as the activation function for the three fully connected layers. As shown in Fig.~\ref{fig1},  the input of the component is one of the set of triplets extracted from a signed social network like $(v_i, v_j, v_k)$ with $e_{ij} = 1$ and $e_{ik}  = -1$. The output of this component is a set of $d_i$-dimension feature vectors like $\{(\mathbf{x_i}, \mathbf{x_j}, \mathbf{x_k})\}$.

{\bf Hash code learning}: This component is designed to learn hash codes of nodes. We use a fully connected layer and a $sgn(\cdot)$ layer to achieve the aim. In particular, the number of neurons of both layers equals to the length of target hash codes, i.e., $d$. The output of the component is the hash codes $(\mathbf{b_i}, \mathbf{b_j}, \mathbf{b_k})$.

{\bf Loss function}: This component measures how well the given triplet supervised information are satisfied by the learned hash codes by computing the triplet loss function.

The triplet loss function is first proposed by \cite{norouzi2012hamming}, which is also called ranking loss. It assumes that the training data $\mathcal{T}$ includes triplets of items $(v_i, v_j, v_k)$, such that the pair $(v_i, v_j)$ is more similar than $(v_i, v_k)$. The goal is to learn hash codes $(\mathbf{b_i}, \mathbf{b_j}, \mathbf{b_k})$ that $\mathbf{b_j}$ is closer to $\mathbf{b_i}$ than $\mathbf{b_k}$ in Hamming distance. The ranking loss can be written as: 
\begin{equation}
\mathcal{L}=\sum_{(v_i, v_j, v_k)\in \mathcal{T}}[\|\mathbf{b_i}-\mathbf{b_j}\|_H - \|\mathbf{b_i} - \mathbf{b_k}\|_H + 1]_+
\label{eq1}
\end{equation}
where $[\alpha]_+ \equiv max(\alpha, 0)$, $\|\mathbf{b_i}- \mathbf{b_j}\|_H$ is the Hamming distance between similar pair, $\|\mathbf{b_i}- \mathbf{b_k}\|_H$ is the Hamming distance between dissimilar pairs. This loss is zero when $\|\mathbf{b_i}- \mathbf{b_j}\|_H$ is at least one bit closer than $\|\mathbf{b_i}- \mathbf{b_k}\|_H$.

%
However, the loss function has three problems for signed network embedding. Firstly, the minimum Hamming distance between similar and dissimilar codes is set as 1 fixedly, which is unreasonable for hash codes of different length. Secondly, the loss function is discrete, which is hard to be optimized. Thirdly, the loss function cannot fit the signed network well.

In order to tackle the first problem, we introduce a hyper-parameter $\delta$, which balances the distance between similar items and dissimilar items. And we have that $\| \mathbf{b_i} - \mathbf{b_j} \|_H \in [0,d]$ and $\| \mathbf{b_i} - \mathbf{b_k} \|_H \in [0,d]$. In order to let the following inequality be valid, $\delta$ should be within the range $[0, d]$, where $d$ is the dimension of hash codes.
\begin{equation}
\| \mathbf{b_i}-\mathbf{b_j} \|_H + \delta \leq \| \mathbf{b_i}-\mathbf{b_k} \|_H 
\label{eq2}
\end{equation}

In order to deal with the second problem, we replace the Hamming distance with the inner production between hash codes inspired by DPSH \cite{li2015feature}. Let $\Theta_{ij}$ denotes half of the inner product between two hash codes $\mathbf{b_i}, \mathbf{b_j} \in \{-1, +1\}^d$:
\begin{equation}
\Theta_{ij} = \frac{1}{2}\mathbf{b_i}^T \mathbf{b_j}
\label{eq3}
\end{equation}

According to Eq.~\eqref{eq3}, we can derive following equation:
\begin{equation}
\| \mathbf{b_i} - \mathbf{b_j} \|_H = \frac{1}{2}(d - 2\Theta_{ij})
\label{eq4}
\end{equation}
where $d$ is the dimension of hash codes. This is the linear relationship between Hamming distance and inner production.

After substitute Eq.~\eqref{eq3} into Eq.~\eqref{eq1}, we can derive that
\begin{equation}
\begin{split}
 \mathcal{L} =&\sum_{(v_i, v_j, v_k)\in \mathcal{T}}[\Theta_{ik} - \Theta_{ij} + \delta]_+ \\
 s.t.&\\
 &\mathbf{b_i}, \mathbf{b_j}, \mathbf{b_k}  \in \{-1, +1 \}^d
\end{split}
\label{eq5}
\end{equation}

Minimizing the above loss function is an intractable discrete optimization problem \cite{li2015feature}, which means we cannot compute the gradient of it. Inspired by LFH \cite{zhang2014supervised}, we relax $\mathcal{B}$ from discrete to continuous, i.e., relax $\mathcal{B}$ to $\mathcal{X}$, where $x_i \in \mathbb{R}^d$. Note that this process will induce relaxation error, which is called quantization error in feature hashing area \cite{wang2016deep}.

Thus, we rewrite the inner production as
\begin{equation}
\Theta_{ij} = \frac{1}{2}\mathbf{x_i}^T \mathbf{x_j}
\label{eq6}
\end{equation}
and the objective function is:
\begin{equation}
\begin{split}
 \mathcal{L} = &\sum_{(v_i, v_j, v_k) \in T}[\Theta_{ik} - \Theta_{ij} + \delta]_+ \\
& + \eta \sum_{m=1}^M \|\mathbf{b_m}-\mathbf{x_m}\|_2^2 
\end{split}
\label{eq7}
\end{equation}
where $\eta$ is the hyper-parameters to balance the triplet loss function and the quantization error. The $\mathbf{b_m} = sgn( \mathbf{x_m} )$, where the $sgn( \cdot )$ is sign function with the threshold zero.

Thirdly, in a real-world signed social network, loss function in Eq.~\eqref{eq7} cannot deal with those nodes whose 2-hop network \cite{wang2017signed} has only positive or negative links. To address this problem, we introduce a virtual node $v_0$, and then create a negative link between $v_0$ and each node whose 2-hop network has only positive links. In this way, we create another training set of triplets $T_0$. Inspired by all the aforementioned methods, the final formulation of hash loss function for signed social network is as follows:
\begin{equation}
\begin{split}
\min_{\mathbf{\mathcal {X}}, \vartheta, \mathbf{x_0}} & \sum_{(v_i, v_j, v_k) \in \mathcal T} [ \Theta_{ik} - \Theta_{ij} + \delta   ]_+ \\
& + \sum_{(v_i, v_j, v_0) \in \mathcal T_0} [ \Theta_{i0} - \Theta_{ij}+ \delta_0 ]_+  \\
& + \alpha \mathcal{R}(\vartheta)+ \eta \sum_{m=1}^M \|\mathbf{b_m}-\mathbf{x_m}\|_2^2 
\end{split}
\label{eq8}
\end{equation}
where $\vartheta$ is a set of parameters of the network feature learning and hash code learning components, $\mathbf{b_0}$ is the hash code of the virtual node $v_0$. $\mathcal{R}(\vartheta)$ is the regularization to avoid the over-fitting, which are based on $l_2$-norm of $\vartheta$. $\alpha$ is a parameter to control the contribution of the regularization.

\subsection{Training detail}
In order to get training triplets from a signed social network, we use positive graph and negative graph to represent the signed social network, where positive graph is a network with only positive links and negative graph is a network with only negative links. Then, a set of parameter $\vartheta$ of our model is updated automatically by the autograd module of Pytorch. Besides, inspired by other deep methods \cite{smith2017cyclical}, \cite{wang2016deep}, we adopt the simplest learning rate schedule, which is to decrease the learning rate linearly from a large initial value to a small value. The detail training stage of the proposed method HSNE is summarized in Algorithm \ref{alg1}. From line 1 to line 10, the training triplets are sampled from a signed social network. In line 11, we initialize the parameters of deep network and train the deep network from line 12 to 19.

\begin{algorithm}[t] 
\caption{Training Stage of HSNE} 
\label{alg:Framwork} 
\begin{algorithmic}[1] 
\REQUIRE Graph: $\mathcal{G} =( \mathcal{V}, \mathcal{E} )$, the dimension of each layer of the framework: $d_i$, the dimension of hash code: $d$, the sign hash function: $sgn(\cdot)$, hyper-parameters: $\delta, \delta_0$ and $\alpha$, the training epochs: t.
\ENSURE Parameters $\vartheta$ for the deep framework and hash codes  $\mathbf{\mathcal{B}}$ for nodes.
\STATE Initialize $\mathcal{T} = \emptyset$ and $\mathcal{T}_0 = \emptyset$
\STATE {\bf for} $v_j$ in positive graph {\bf do}
\STATE ~~~~{\bf for} $v_j$ neighbors $v_i$ in positive graph {\bf do}
\STATE ~~~~~~~~{\bf if} $v_i$ in negative graph:
\STATE ~~~~~~~~~~~~$v_k$ is neighbors of $v_i$ in positive graph
\STATE ~~~~~~~~~~~~Put $(v_i, v_j, v_k)$ in $\mathcal{T}$
\STATE ~~~~~~~~{\bf else} :
\STATE ~~~~~~~~~~~~Put $(v_i, v_j, v_0)$ in $\mathcal{T}_0$
\STATE ~~~~{\bf end for}
\STATE {\bf end for}
\STATE Initialize parameters $\vartheta$ for the framework
\STATE {\bf for} each epoch {\bf do}
\STATE ~~~~{\bf for} each mini-batch {\bf do}
\STATE ~~~~~~~~Calculate the outputs by forward propagating through the framework.
\STATE ~~~~~~~~Compute the triplet loss and quantization error using Eq.~\eqref{eq8}.
\STATE ~~~~~~~~Update parameters of the framework by back propa-gating.
\STATE ~~~~{\bf end for}
\STATE ~~~~Adjust learning rate
\STATE {\bf end for}
\end{algorithmic}
\label{alg1}
\end{algorithm}

 \section{Time Complexity Analysis}
\textbf{Training:} 
Let $d_0$ be the dimension of the embedding layer; $d_i$ be the number of nodes in the i-th fully connected layer, $1 \leq i \leq 3$; $d$ be the dimension of hash codes. For a triplet, the computational cost of forward and backward propagation between two layers is $\mathcal{O}(d_1d_2)$. Thus, the time complexity of training this model is $\mathcal{O}(tN(\sum_{i=1}^3d_{i-1}di+d_3 d))$, where $t$ is training epochs and $N$ is the size of training dataset.

\textbf{Similarity search in embedding space:} 
Many processing and analysis tasks over a signed social network, such as link prediction and node classification, are approximate nearest neighbor (ANN) search tasks in embedding space. As the scale of network growing larger and larger, ANN search inside such embedding space becomes impractical using traditional approaches due to their high time complexity, which are typically $\mathcal{O}(n^2)$. Hashing approaches can significantly reduce time and space consumption to an acceptable level with a small and controllable sacrifice of performance as trade-off. For example, a survey \cite{wang2014hashing} shows that a typical time complexity is $\mathcal{O}(n)$ for hashing similarity search, and the space consumption is known to be small, due to the nature of hash codes. Thus, our method uses deep hashing to address this problem.

\section{Experiments}

In this section, We first introduce two benchmark datasets, and then discuss the results of our method as well as several state-of-the-art methods over link prediction task. Finally, we evaluate the sensitivity of parameters, and test the parameters' effect on the quality of node embedding. 

\subsection{Datasets}
Two signed social networks, Epinions \cite{wang2017signed} and Slashdot \cite{wang2017signed}, \cite{yuan2017sne}, are used as benchmark datasets in our experiments. Epinions is a dense signed social network in which there are 131,828 users and 841,372 relationships, of which about 85.3\% are trust relationships. The reviewers connected with positive and negative links express the trust and distrust relationships. Slashdot is a sparse network of friends, which allows users to tag each other as ``friends'' (like) or ``foes'' (dislike). The dataset is comprised of 77,357 users and 516,575 relationships of which 76.7\% are ``friend'' relationships. The dataset can be used to infer the ``friend'' relationships between users, and to study the positive and negative influence. Both networks are directed, and we filter out the repeated links and users who have no links. Some key statistics of two datasets are summarized in Table \ref{tab1}.

\begin{table}[htbp!]
\caption{Statistics of Epinions and Slashdot }
\begin{center}
\scalebox{0.9}{
\begin{tabular}{ | c | c | c | }
	\hline
	 &  Epinions (Dense) &  Slashdot (Sparse) \\ \hline
    Type            & Directed & Directed \\ \hline	
    \# of Users     & 131,828 & 77,350 \\ \hline
    \# of Links (+) & 717,667 & 396,378 \\ \hline
    \# of Links (-) & 123,705 & 120,197 \\ \hline
\end{tabular}}
\label{tab1}
\end{center}
\end{table}

\subsection{Link Prediction}
Link prediction, as one of the most fundamental problems in network analysis, has received a considerable amount of attention \cite{liben2007link}, \cite{lu2011link}. It aims to estimate the likelihood of the existence of an edge between two nodes based on observed network structure \cite{getoor2005link}. This paper will take the link prediction as testing task to evaluate the performance over two real-world signed social network, i.e., Epinions and Slashdot.

Following the work \cite{grover2016node2vec}, we regard link prediction as a classification task. Specifically,  we first use node representation to compose link representations with the following operators: (1) $hadamard$; (2) $average$; (3) $l1\_weight$ and (4) $l2\_weight$. Based on the link representation, a one-vs-rest logistic regression classifier is trained  by 10-fold cross validation to predict whether there is a positive or negative links between two nodes. As is shown in Table \ref{tab1}, positive links are much denser than negative links in both signed social networks. Therefore, we use the average AUC to evaluate the link prediction problem instead of accuracy \cite{leskovec2010predicting}, \cite{wang2017signed}. The baselines are as follows:  

\begin{itemize}
\item Node2Hash \cite{wang2018feature} is an feature hashing based method. It uses the encoder-decoder framework, where the encoder is used to map the proximity of nodes into a feature space, and the decoder is used to generate node embedding through feature hashing.
\item DeepWalk \cite{perozzi2014deepwalk} is an unsupervised method, it uses local information obtained from truncated random walks to learn dimension feature representations of nodes.
\item Line \cite{tang2015line} uses the breadth-first strategy to sample the inputs, based on node neighbors.The method preserves both the first order and second order proximities in node embedding process.
\item SiNE \cite{wang2017signed} is a deep learning framework designed for signed social network. The framework optimize an objective function guided by social theories.
\item SNE \cite{yuan2017sne} adopts the log-bilinear model to obtain node representations of all nodes along a given path. Moreover, it incorporates two signed-typed vectors to capture the positive or negative relationship of each edge along the path.
\end{itemize}

Note that there are two types of baselines, one is hashing based method, i.e., Node2Hash; the other is non-hashing methods. Generally, hashing based methods performs worse than non-hashing methods with higher efficiency. In this paper, the proposed method is a kind of hashing based method, thus our aim is to exceed the baseline Node2Hash, not other non-hashing methods. In addition, to the best of our known, there only have two hashing based methods. NetHash \cite{wu2018efficient} is the another one, but cannot be chosen as our baseline because the method needs attribute information.  

\begin{table}[b]
\caption{Comparing the AUC for signed link prediction between HSNE and hashing based method on Epinions and Slashdot}
\begin{center}
\scalebox{0.85}{\begin{tabular}{ | c | c | c | c | c | c |}
	\hline
	Dataset & Approach  & $hadamard$        & $average$         & $l1$              & $l2$  \\
	\hline
\multirow{2}{*}{\shortstack{Epinions\\(Dense)}}
    & Node2Hash & 0.7353          & 0.6600          & 0.6145          & 0.6640         \\
    \cline{2-6}
    & HSNE      & \textbf{0.8145}          & \textbf{0.7675}          & \textbf{0.8144}          & \textbf{0.8144} \\
    \cline{1-6} 
\multirow{2}{*}{\shortstack{Slashdot\\(Sparse)}}
    & Node2Hash & 0.7798          & 0.6470           & 0.6112          & 0.6364         \\
    \cline{2-6} 
    & HSNE      & \textbf{0.7808}          & \textbf{0.7439}           & \textbf{0.7807}          & \textbf{0.7807}         \\
    \hline 
\end{tabular}}
\label{tab2}
\end{center}
\end{table}

\subsection{Results and Discussion}
For proposed HSNE, we empirically set the size of deep neural network $d_0 = 200$, $d_i = 320$ and $d = 256$, which means the dimension of embedding layer is 200, the size of hidden layer dimension is 320 and hash code dimension is 256. Besides, for Epinions, we set the hyper-parameters learning rate $lr = 0.009$, $\delta = 24$, $\delta_0 = 12$ and $\eta = 40$. Slashdot is a sparse network, comparing to Epinions. Thus, the hyper-parameters will be changed, and are set as that $\delta = 16$, $\delta_0 = 8$ and $\eta = 0.55$. 

In this subsection, we compare our HSNE with hashing based network embedding method, and average AUC are reported in Table \ref{tab2}. Besides, we also compare our HSNE with non-hashing based network embedding methods in the Table \ref{tab3} and Table \ref{tab4}, where type "P" expresses that the methods only consider positive links and type "P \& N" expresses that the methods consider both positive and negative links.

According to the Table \ref{tab2}, we can obtain the following observation:
\begin{itemize}
\item The performance of the propose method HSNE has a significant improvement over Node2Hash in all the operators. Node2Hash only considers the positive links and ignores negative links. On the contrary, Our method considers simultaneously positive and negative links, which reduces the loss of network information. The result suggests that our method can improve the performance of hashing based signed social network embedding.

\item HSNE greatly outperforms Node2Hash in $l1\_weight$ and $l2\_weight$ operators. The main reason is that the two operators are similar to Hamming distance. According to the objective function Eq.~\ref{eq8}, HSNE uses Hamming distance to express the similarity of two nodes. Thus, the two operators can obtain much better performance.
\end{itemize}

According to the Table \ref{tab3} and Table \ref{tab4}, we can obtain the following observations:
\begin{itemize}
\item Generally, binary representation leads to accuracy loss. However, comparing to other non-hashing methods, HSNE can still achieve the near-best or even-best results with higher efficiency in some cases. For example, we compare HSNE with type "P" methods in $l1\_weight$ and $l2\_weight$ operators in Table \ref{tab3} and all the other methods in $average$ operator in Table \ref{tab4}. Thus, our method is still extremely competitive.

\item The effectiveness of "P \& N" type methods are significantly better than "P" type ones in most cases, which also prove that negative links can help improve the performance. For example, although both of SNE and DeepWalk adopt Random Walk to sample features from the same signed social network, SNE is better than DeepWalk over all the operators.
\end{itemize}

\subsection{Parameter Analysis}
In this subsection we investigate the impact of hyper-parameters learning rate $lr$, $\delta$, $\delta_0$, $\eta$ and the number of  fully connected layers $L$ on the performance of link prediction. To investigate the impact and sensitivity of HSNE on those hyper-parameters, we fix epoch = 100 and the size of network $d_0=200$, $d_i = 320$ and $d = 256$. In addition, we use the $hadamard$ operator to compose link representations and compute AUC of link prediction task.

\begin{table}[t!]
\caption{Comparing the AUC for signed link prediction between HSNE and non-hashing based methods on Epinions}
\begin{center}
\scalebox{0.9}{\begin{tabular}{ | c | c | c | c | c | c |}
	\hline
	Type & Approach  & $hadamard$        & $average$         & $l1$              & $l2$  \\
	\hline
\multirow{2}{*}{P}
    & DeepWalk  & 0.9159          & 0.7831          & 0.7566          & 0.7280         \\ 
    \cline{2-6}
    & Line      & 0.9011          & 0.7835          & 0.7814          & 0.7785         \\
    \cline{1-6}                         
\multirow{3}{*}{P \& N}
    & SiNE      & 0.9239          & 0.8032          & \textbf{0.9016}          & 0.8932         \\
    \cline{2-6} 
    & SNE       & \textbf{0.9524}          & \textbf{0.8386}          & 0.8967          & \textbf{0.9119}         \\
    \cline{2-6} 
    & HSNE      & 0.8145          & 0.7675          & 0.8144          & 0.8144         \\
    \hline 
\end{tabular}}
\label{tab3}
\end{center}
\end{table}

\begin{table}[t!]
\caption{Comparing the AUC for signed link prediction between HSNE and non-hashing based methods on Slashdot}
\begin{center}
\scalebox{0.9}{\begin{tabular}{ | c | c | c | c | c | c |}
	\hline
	Type & Approach  & $hadamard$        & $average$         & $l1$              & $l2$  \\
	\hline
\multirow{2}*{P}
    & DeepWalk  & 0.8226          & 0.6918           & 0.7163          & 0.7143         \\ 
    \cline{2-6}
    & Line      & 0.8876          & 0.7006          & 0.6500          & 0.6489         \\
    \cline{1-6}                             
\multirow{3}*{P \& N}
    & SiNE      & 0.8963          & 0.6432           & \textbf{0.7983}          & 0.8031         \\
    \cline{2-6}
    & SNE       & \textbf{0.9017}          & 0.7043           & 0.7801          & \textbf{0.8198}         \\
    \cline{2-6}
    & HSNE      & 0.7808          & \textbf{0.7439}           & 0.7807          & 0.7807         \\
    \hline
\end{tabular}}
\label{tab4}
\end{center}
\end{table}

\begin{figure}[t!]
    \centering
    \includegraphics[width=8cm,height=5cm]{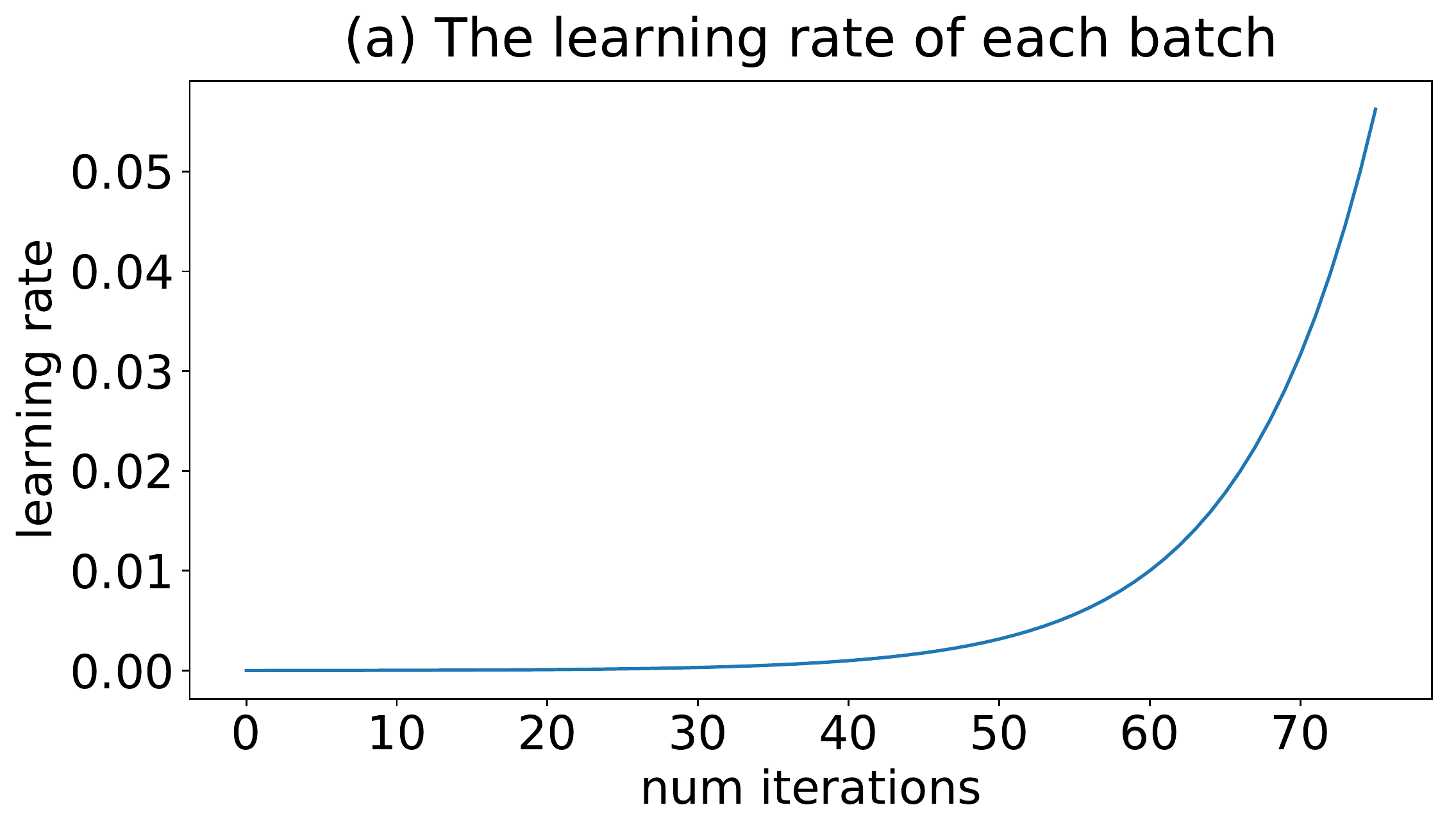}
    \includegraphics[width=8cm,height=5cm]{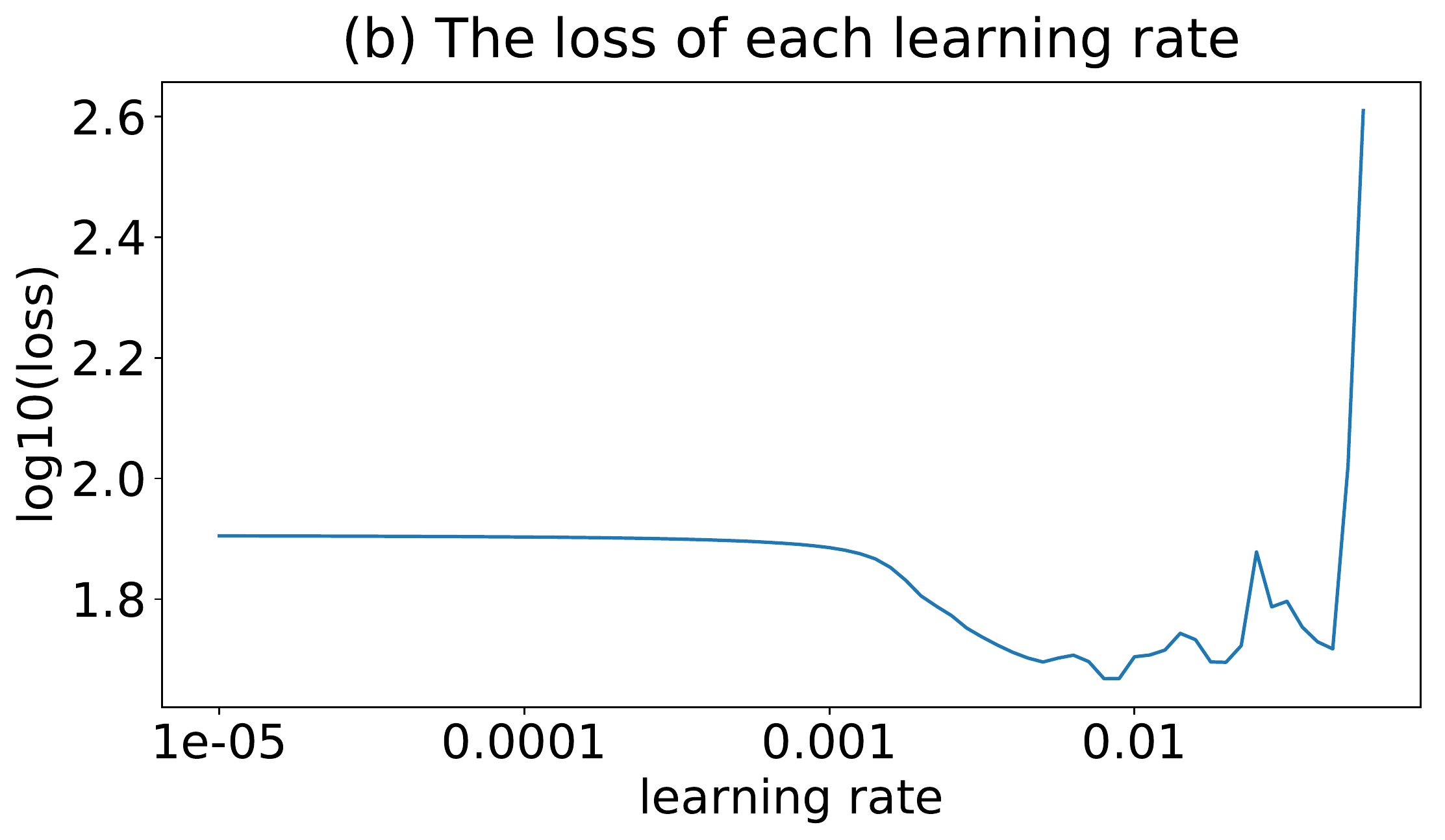}
    \caption{Impact of the initial learning rate $lr$}
    \label{fig3}
    \end{figure}

\textbf{Impact of the initial learning rate $lr$:} The learning rate is the most important hyper-parameter to tune for deep neural networks, which is to balance the loss and the training speed. In order to select a proper initial learning rate over Epinions, we fix $\delta = 24$, $\delta_0 = 12$ and $\eta = 40$. Inspired by the work \cite{smith2017cyclical}, we increase the learning rate from the minimum value 1e-5 to the maximum value 1, and update parameters of network after each batch over Epinions. At the same time, we record the loss of each batch. Fig.~\ref{fig3} (a) shows the learning rate of each batch and Fig.~\ref{fig3} (b) shows the loss of each learning rate. The experimental results show that a value of the initial learning rate around 0.009 gives relatively good loss. 

\begin{figure}[t!]
    \centering
    \includegraphics[width=8cm,height=5cm]{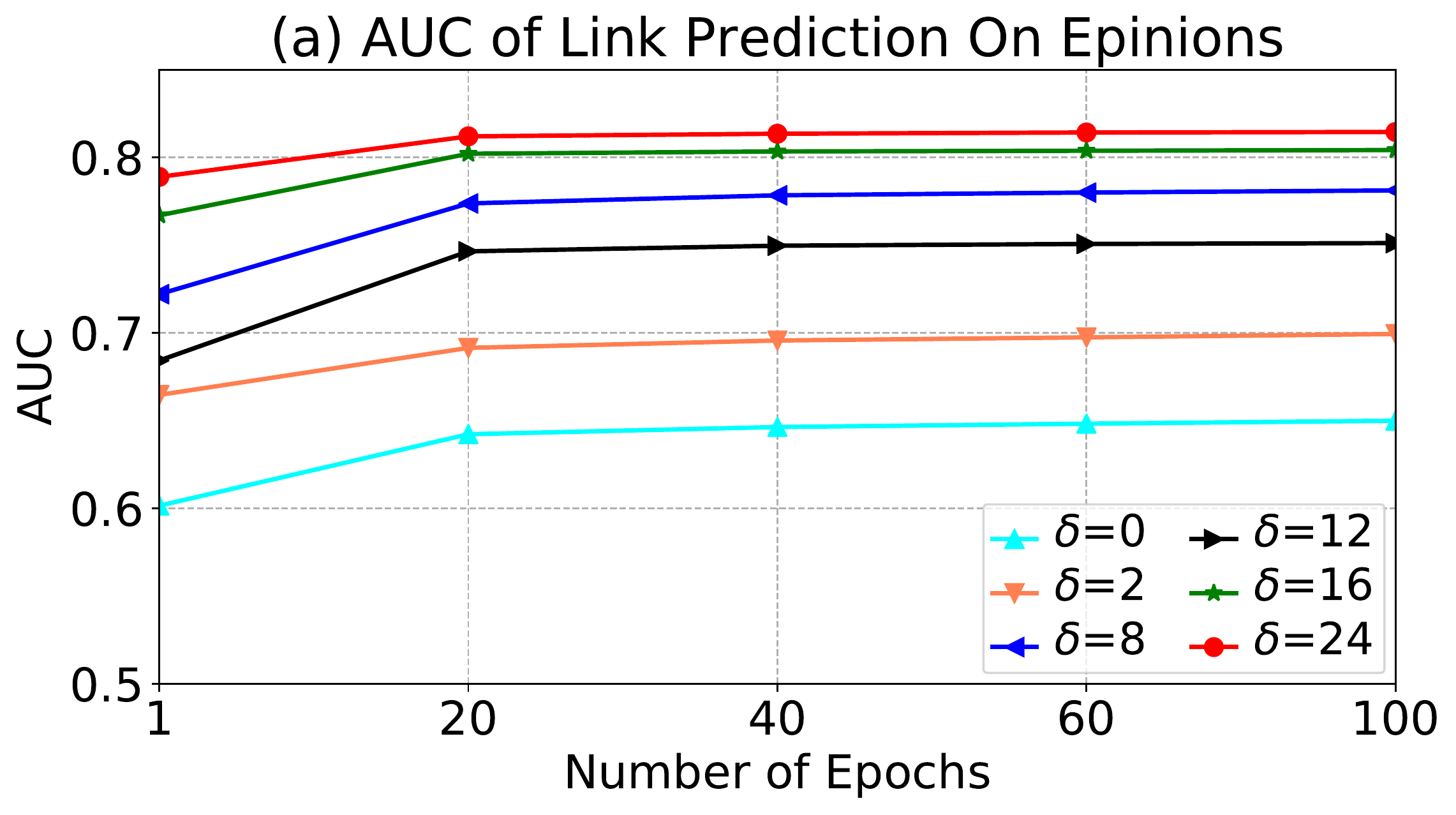}
    \includegraphics[width=8cm,height=5cm]{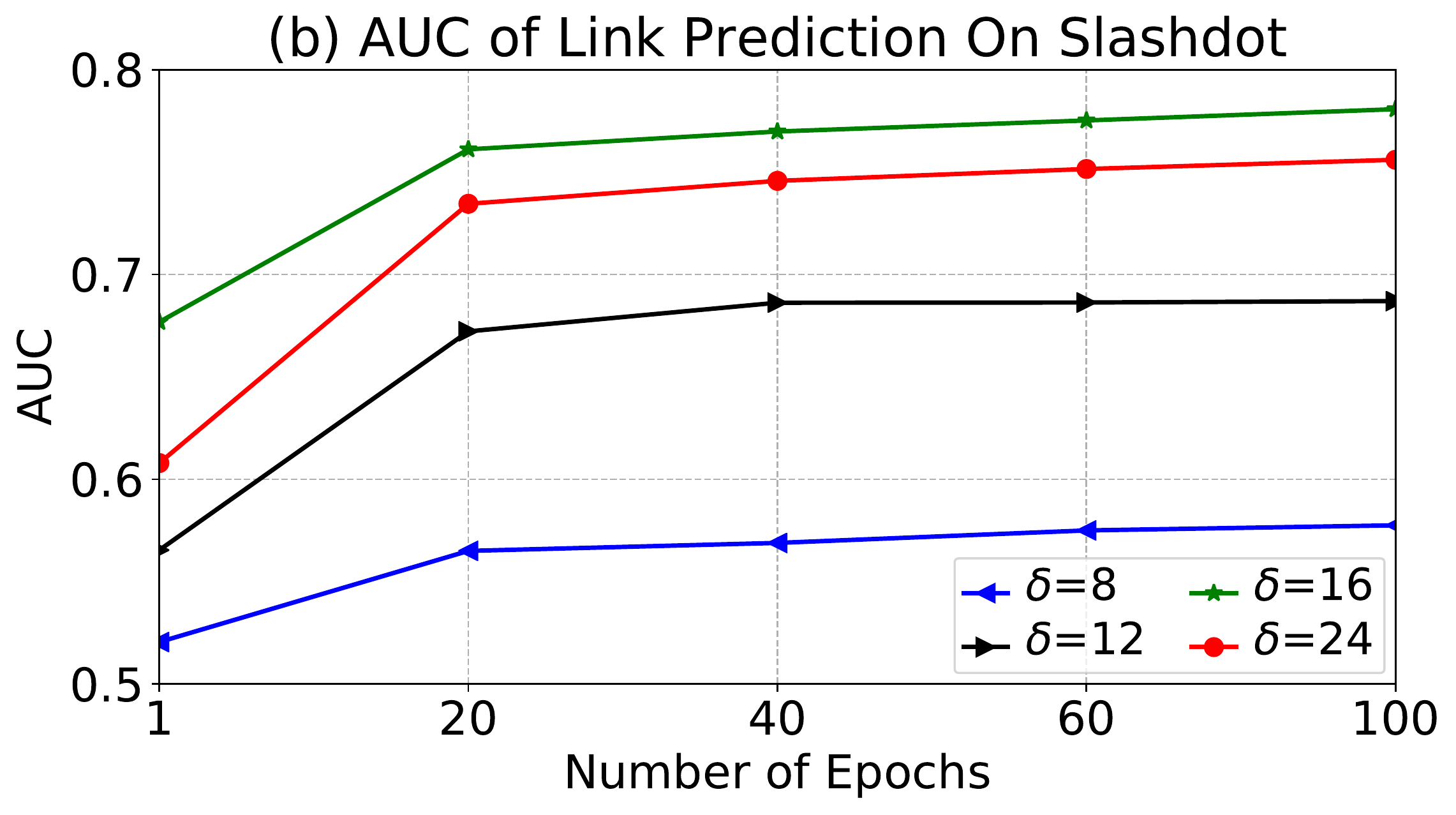}
    \caption{Impact of hyper-parameter $\delta$ and $\delta_0$}
    \label{fig4}
\end{figure}

\textbf{Impact of hyper-parameter $\delta$ and $\delta_0$:} As shown in Eq.~\eqref{eq2}, hyper-parameter $\delta$ can balance the distance between similar items and dissimilar items. If $\delta = 0$, $\| \mathbf{b_i}-\mathbf{b_k} \|_H - \| \mathbf{b_i}-\mathbf{b_j} \|_H$  will be very close to zero which means the model fail to distinguish similarity and dissimilarity. However, if the value of $\delta$ is too large, it will break the balance of the triplet loss function and the quantization error. According to the work \cite{wang2017signed}, $\delta_0$ is $\frac{1}{2} \delta$, which has the same effect as $\delta$. To investigate the effects of $\delta$ and $\delta_0$, we fix $\eta = 40$ (0.55) over Epinions (Slashdot), the initial learning rate $lr = 0.009$ and $\delta = 2\delta_0$. Fig.~\ref{fig4} shows that a big value may gives relatively good AUC. Besides, $\delta$ and $\delta_0$ is more sensitive over sparse networks, i.e., Slashdot. 

\begin{figure}[t!]
    \centering
    \includegraphics[width=8cm,height=5cm]{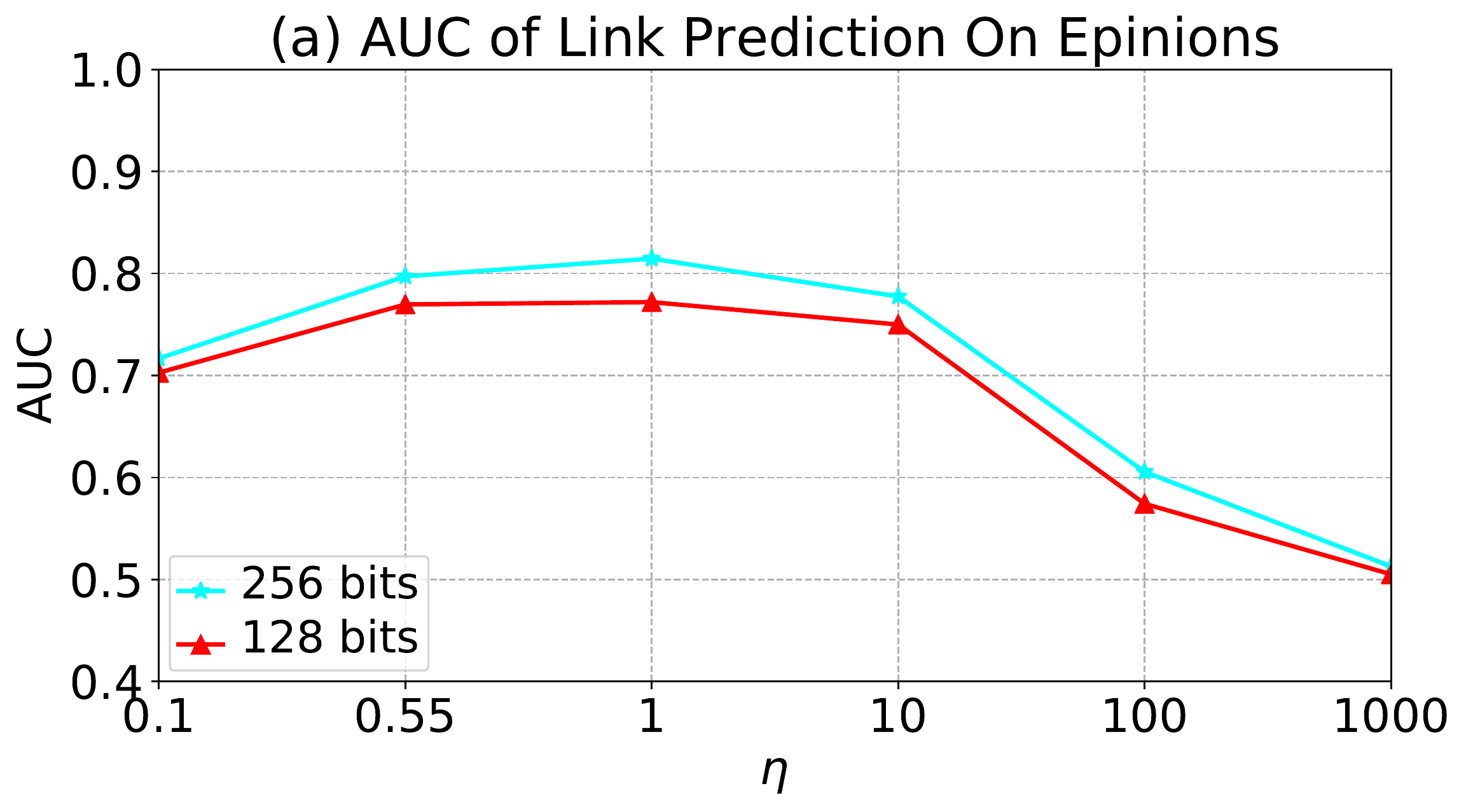}
    \includegraphics[width=8cm,height=5cm]{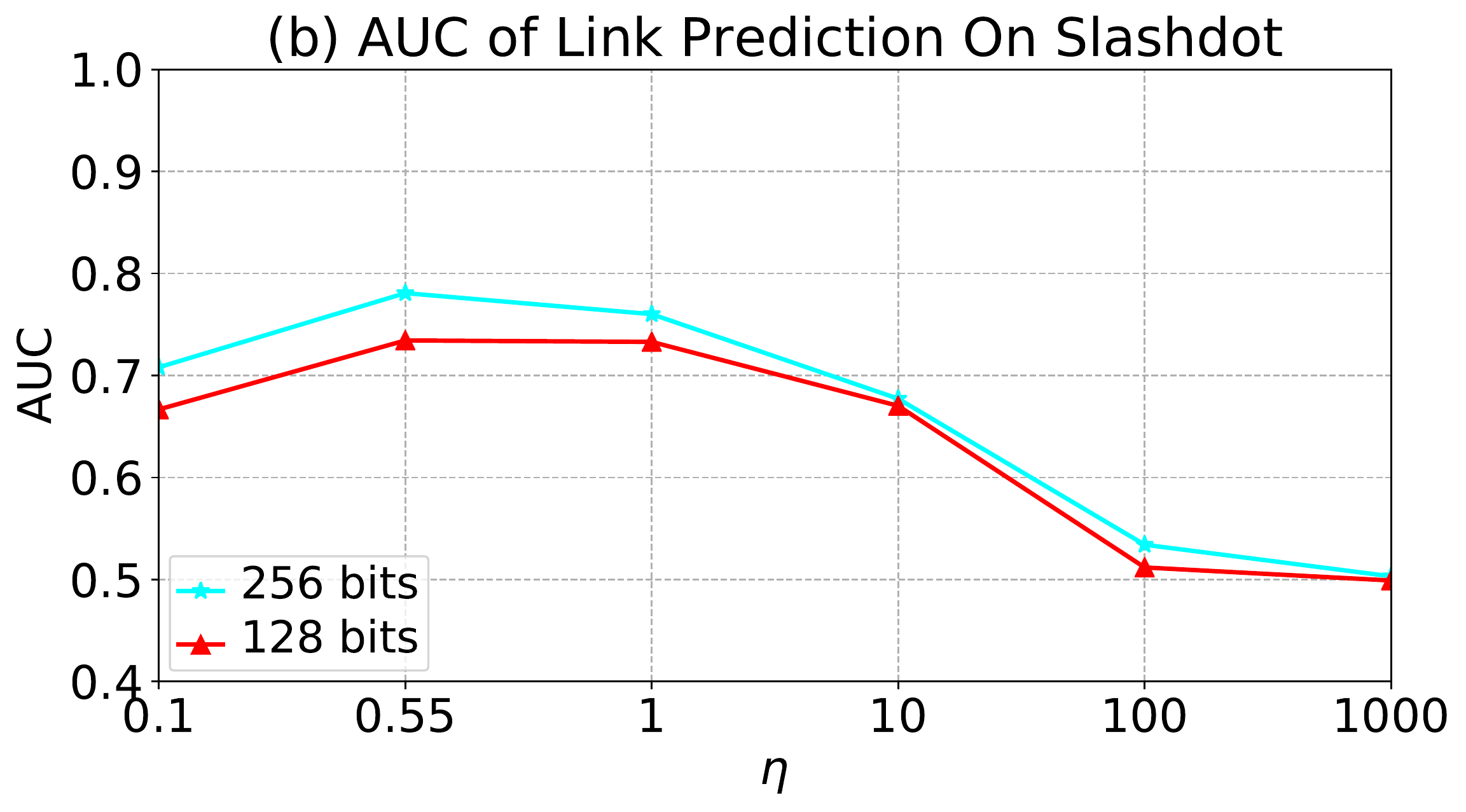}
    \caption{Impact of hyper-parameter $\eta$}
    \label{fig5}
    \end{figure}
    
    \begin{figure}[t!]
        \centering
        \includegraphics[width=8cm,height=5cm]{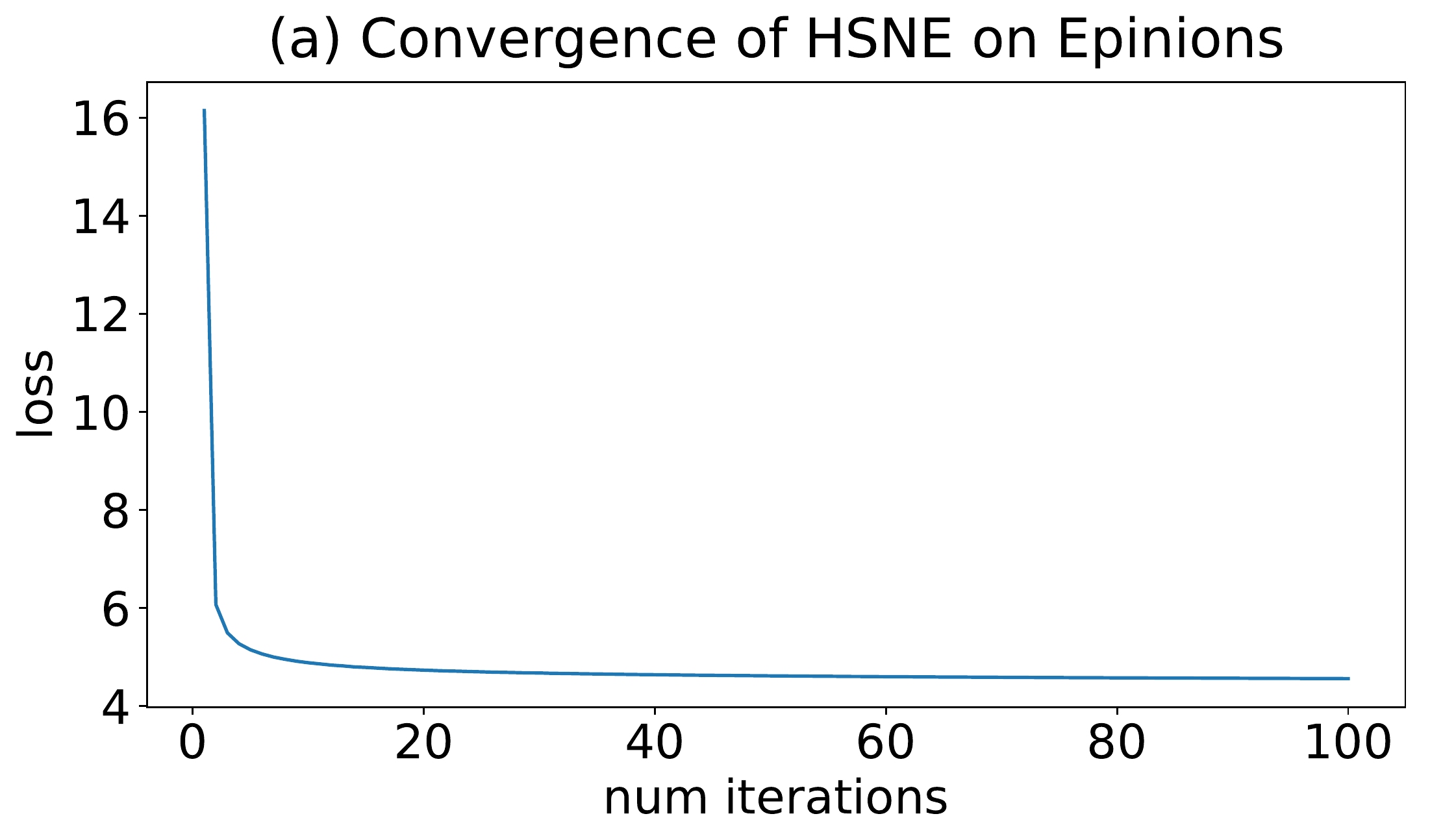}
        \includegraphics[width=8cm,height=5cm]{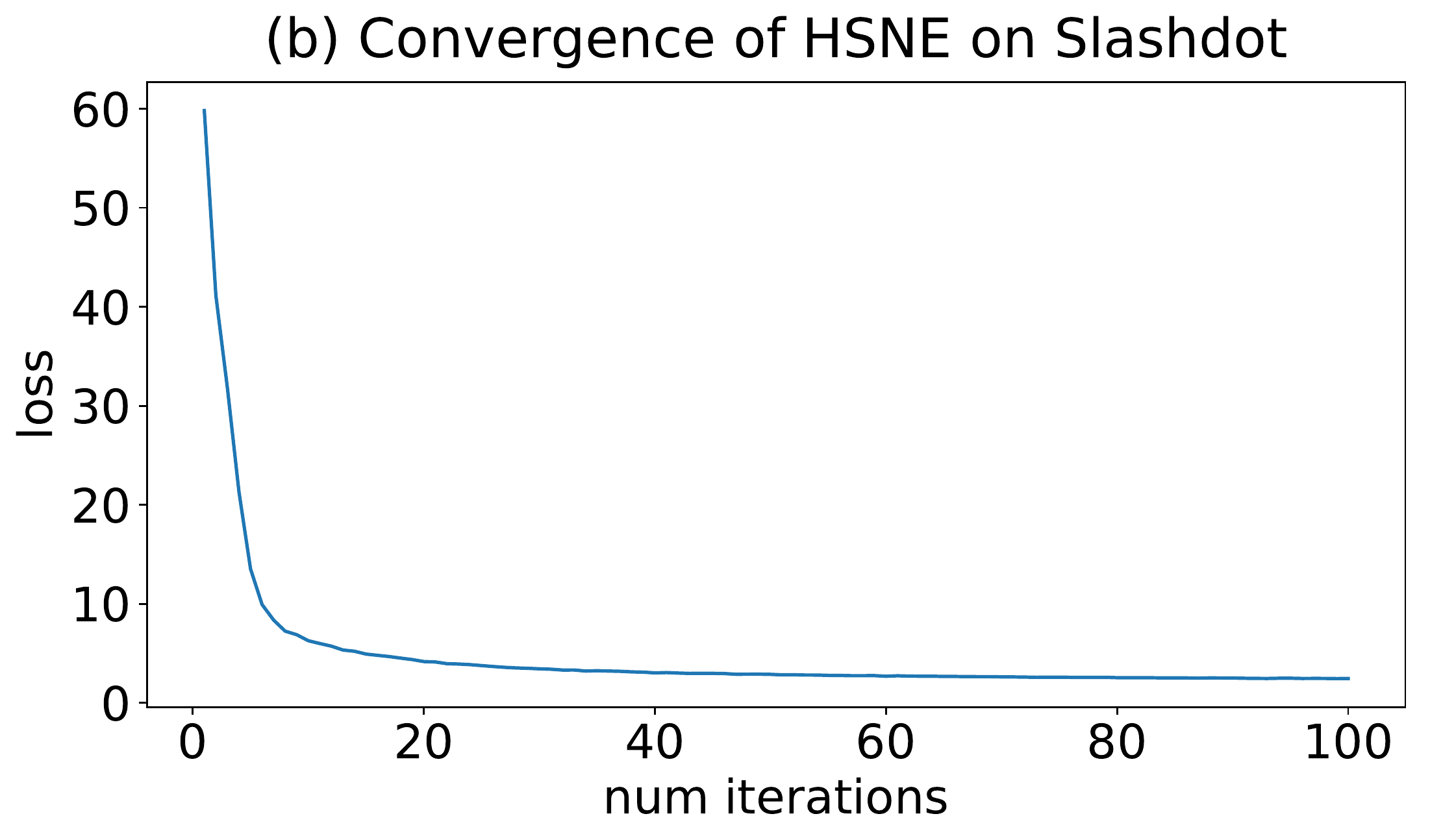}
        \caption{Convergence of HSNE on Epinions and Slashdot}
        \label{fig2}
\end{figure}

\textbf{Impact of hyper-parameter $\eta$:} As shown in Eq.~\eqref{eq7}, $\eta$ is designed to balance the triplet loss function and the quantization error. To investigate the effects of $\eta$, we fix $\delta = 24$ (16) over Epinions (Slashdot), $\delta_0 = \frac{1}{2} \delta$ and the initial learning rate $lr = 0.009$. As shown in Fig.~\ref{fig5},  there is a significant performance drop in terms of AUC when $\eta$ becomes very small (e.g., 0.1) or very large (e.g., 1000). It is reasonable since it is designed to balance the triplet loss function and the quantization error. Thus, it will lead to imbalance between these two terms, if the value of $\eta$ is very small or very large. Besides, $\eta$ is more sensitive over sparse network, i.e., Slashdot.

\textbf{Impact of $L$:}  $L$ is the number of fully connected layers in network feature learning component. To investigate the impact of L, we fix $\eta = 40$ (0.55), $\delta = 24$ (16) over Epinions (Slashdot), $\delta_0 = \frac{1}{2} \delta$ and the initial learning rate $lr = 0.009$. Then, we vary L as 1, 2, 3, 4, 5 and record results. The final results in term of AUC are reported in Table \ref{tab5}. According to the Table \ref{tab5}, we can obtain the following observation that as the increase of $L$, the performance increases first and then decreases, which suggests that we can learn a relatively good embedding with L=3. According to the loss, we believe, the main reason for the decrease is that as the network becomes deeper, its harder to train.

\begin{table}[t!]
    \caption{AUC of HSNE on signed link prediction with different number of fully connected layers $L$}
    \begin{center}
    \scalebox{0.9}{\begin{tabular}{ | c | c | c | c | c | c |}
        \hline
        Dataset     & $L=1$     & $L=2$     & $L=3$     & $L=4$     & $L=5$  \\
        \hline
        Epinions  & 0.7122    & 0.7444    & \textbf{0.8145}    & 0.7566    & 0.7421     \\
        \hline
        Slashdot  & 0.6799    & 0.7241    & \textbf{0.7808}    & 0.7521    & 0.7361     \\
        \hline
    \end{tabular}}
    \label{tab5}
    \end{center}
\end{table}

\subsection{Convergence Analysis}
According to Eq.~\eqref{eq8}, the objective function of HSNE is non-negative. Thus, by minimizing it using gradient descent and choosing a proper initial learning rate, the objective function will achieve an optimal point or a local optimal point and converge. In this subsection, We plot the value of the objective function in each training epoch for both Epinions and Slashdot in Fig.~\ref{fig2}. From the figure, we can see that the value of the objective function decreases fast at the first 20 iterations and then gradually converges. For both datasets, it takes about 60 epochs to converge.

\section{Conclusion}
The existing feature hashing based network embedding algorithms only consider the positive links and ignore negative links in signed social networks. To address this problem, we propose a novel deep hashing based network embedding method by considering simultaneously positive and negative links in a signed social network. Experimental results on link prediction over two real-world signed social network show that our method outperforms the state-of-the-art hashing based network embedding approach, which only considers positive links.

\bibliography{fixbib}
\bibliographystyle{aaai}
\end{document}